\documentstyle[11pt,aaspp4]{article}

\received{}
\revised{}
\accepted{}
\journalid{}{}
\articleid{}{}
\paperid{}
\cpright{}{}
\ccc{}

\lefthead{Wood \& Loeb}
\righthead{Escape of Lyman Continuum Radiation}

\begin{document}

\title{Escape of Ionizing Radiation from High--Redshift Galaxies}

\author{Kenneth Wood\altaffilmark{1} and Abraham Loeb\altaffilmark{2}}
\affil{Harvard-Smithsonian Center for Astrophysics,
60 Garden Street, Cambridge, MA 02138}

\altaffiltext{1}{kenny@claymore.harvard.edu}
\altaffiltext{2}{aloeb@cfa.harvard.edu}

\authoremail{kenny@claymore.harvard.edu}

\begin{abstract}
We use a three-dimensional radiation transfer code to calculate the
steady-state escape fraction of ionizing photons from disk galaxies as a
function of redshift and galaxy mass. The gaseous disks are assumed to be
isothermal (with a sound speed $c_s\sim 10~{\rm km~s^{-1}}$) and radially
exponential. Their scale-radius is related to the characteristic spin
parameter and virial radius of their host halos, and their vertical
structure is dictated by their self-gravity. The sources of radiation are
taken to be either stars embedded in the disk, or a central quasar. The
predicted increase in the disk density with redshift results in an overall
decline of the escape fraction with increasing redshift. For typical
parameters of smooth disks, we find that the escape fraction at $z\sim 10$
is $\la 1\%$ for stars, but $\ga 30\%$ for mini-quasars.  Unless the smooth
gas content of high-redshift disks was depleted by more than an order of
magnitude due to supernovae-driven outflows or fragmentation, the
reionization of the universe was most likely dominated by mini-quasars
rather than by stars.

\end{abstract}


\section{Introduction}

Recently, there has been considerable theoretical interest in calculating
the reionization history of the intergalactic medium (e.g., Haiman \& Loeb
1997; Abel, Norman, \& Madau 1999; Madau, Haardt, \& Rees 1999;
Miralda-Escud\'e, Haehnelt, \& Rees 1999; Gnedin 1999).  The intergalactic
ionizing radiation field is an essential ingredient in these calculations
and is determined by the amount of ionizing radiation escaping from the
host galaxies of stars and quasars.  The value of the escape fraction as a
function of redshift and galaxy mass remains a major uncertainty in all
current studies, and could affect the cumulative radiation intensity by
orders of magnitude at any given redshift.  In general, the gas density
increases towards the location of galaxies in the intergalactic medium
(IGM) and so the transfer of the ionizing radiation must be followed
through the densest regions on the galactic length scales.  Reionization
simulations are limited in dynamical range and small-scale resolution and
often treat the sources of ionizing radiation (quasars and galaxies) as
unresolved point sources within the large-scale intergalactic medium (see,
e.g.,  simulations by Gnedin 1999).  In this paper we
calculate the escape of ionizing photons from disk galaxies as a function
of formation redshift and mass, thereby providing a means of estimating the
ionizing luminosity input for simulations of reionization.

The escape of ionizing radiation ($h\nu > 13.6$eV, $\lambda < 912${\AA})
from the disks of present-day galaxies has been studied in recent years in
the context of explaining the extensive diffuse ionized layers observed
above the disk in the Milky Way (Reynolds et al. 1995) and other galaxies
(e.g., Rand 1996; Hoopes, Walterbos, \& Rand 1999).  Theoretical models
predict that of order 3--14\% of the ionizing luminosity from O and B stars
escapes the Milky Way disk (Dove \& Shull 1994; Dove, Shull, \& Ferrara
1999).  A similar escape fraction of $f_{\rm esc}=6$\% was determined by
Bland-Hawthorn \& Maloney (1998) based on H$\alpha$ measurements of the
Magellanic Stream.  From {\it Hopkins Ultraviolet Telescope} observations
of four nearby starburst galaxies (Leitherer et al. 1995; Hurwitz,
Jelinsky, \& Dixon 1997), the escape fraction was estimated to be in the
range 3\%$<f_{\rm esc} < 57$\%.  If similar escape fractions characterize
high redshift galaxies, then stars could have provided a major fraction of
the background radiation that reionized the IGM (Madau \& Shull 1996; Madau
1999).  However, the escape fraction from high-redshift galaxies, which
formed when the universe was much denser ($\rho\propto (1+z)^3$), may be
significantly lower than that predicted by models that are adequate for
present-day galaxies.

In popular Cold Dark Matter (CDM) models, the first stars and quasars
formed at redshifts $z\ga10$ (see, e.g. Haiman \& Loeb 1997, 1998; Gnedin
1999).  These sources are expected to have formed in galaxies where the gas
has cooled significantly below its virial temperature and has assembled
into a rotationally-supported disk configuration (Barkana \& Loeb 1999a,b).
The ionizing radiation leaving their interstellar environments, and
ultimately their host galaxies, streamed into intergalactic space creating
localized ionized regions.  Through time these ionized regions, or
``Str\"omgren volumes,'' expanded and eventually overlapped at the epoch of
reionization (Arons \& Wingert 1972).  The redshift of reionization is
still unknown, but the detection of flux shortward of the Ly$\alpha$
resonance for galaxies out to redshifts $z\sim$ 5--6 (Stern et al. 2000;
Weymann et al. 1998; Dey et al. 1998; Spinrad et al. 1998; Hu et al. 1998,
1999), implies that reionization occurred at even higher redshifts (Gunn \&
Peterson 1965).

Current reionization models assume that galaxies are isotropic point
sources of ionizing radiation and adopt escape fractions in the range $5\%
< f_{\rm esc} < 60\%$ (see, e.g., Gnedin 1999, Miralda-Escud\'e et
al. 1999).  In this paper, we examine the validity of these assumptions by
following the radiation transfer of ionizing photons in the gaseous disks
of high redshift galaxies. We consider either stars or a central quasar as
the sources of ionizing photons.  The mass and radial extent of the gaseous
galactic disks, in which these sources are embedded, are functions of
redshift and can be related to the mass and radius of their host dark
matter halos (Navarro, Frenk, \& White 1997; Mo, Mao, \& White 1998).  The
vertical structure of the disk is dictated by its self-gravity and will be
assumed to follow the isothermal profile (Spitzer 1942) with a thermal (or
turbulent) speed of $\sim 10~{\rm km~s^{-1}}$ for all galaxies. This
corresponds to the characteristic thermal speed of photoionized gas and
also to an effective gas temperature of $\sim 10^4~{\rm K}$, below which
atomic cooling is suppressed (Binney \& Tremaine 1987).  We relate the
ionizing luminosity emitted by the stars or quasars to the mass of the host
dark matter halos (Haiman \& Loeb 1997, 1998), adopting current estimates
for Lyman continuum production in starburst galaxies (Leitherer et
al. 1999) and quasars (Laor \& Draine 1993).  The escape fractions for disk
galaxies as a function of mass and redshift are then calculated with a
Monte Carlo radiation transfer code, similar to that of Och, Lucy, \& Rosa
(1998).

Our numerical code finds the ionized fraction of the gas and follows the
associated radiation transfer of the ionizing photons in a steady
state. Strictly speaking, it provides an upper limit on the degree of
ionization and the corresponding escape fraction for a given source
luminosity.  However, the characteristic propagation time of the ionization
front through the scale-height of the galactic disks of interest here ($\ll
10^6~{\rm yr}$), is much shorter than the expected decay time of the
starburst or quasar activities which produce the ionizing photons.

For the sake of simplicity, we begin our study of the problem with the
assumption that the gas is distributed smoothly within the disk.  Clumping
is known to have a significant effect on the penetration and escape of
radiation from an inhomogeneous medium (e.g., Boiss\'e 1990; Witt \& Gordon
1996, 2000; Neufeld 1991; Haiman \& Spaans 1999; Bianchi et al. 2000).  The
inclusion of clumpiness will introduce several unknown free parameters into
the calculation, such as the number and density contrast of the clumps, and
the spatial correlation between the clumps and the ionizing sources.  An
additional complication might arise from hydrodynamic feedback, whereby
part of the gas mass might be expelled from the disk by stellar winds and
supernovae (Couchman \& Rees 1986; Dekel \& Silk 1986).  We adopt a simple
approach to gauge the significance of clumpiness in Section~4 by modeling
the galactic disk density as a two-phase medium (Witt \& Gordon 1996,
2000) and calculate the escape fractions from such clumpy disks.  We also
simulate the effects of outflows and the possible expulsion of gas from the
disk by reducing the mass of the smoothly distributed gas in the disk by an
order of magnitude.

In \S~2 we present the details of our model, including the disk geometry,
the gas distribution, the source luminosities, and the radiation transfer
code. \S~3 presents the results of our models in terms of the derived
escape fractions. Our investigation into clumpy disks is presented in
\S~4. We conclude with a discussion of our findings in \S~5.

\section{Model Ingredients}

In order to determine the escape of ionizing photons we need to specify the
geometry and density of the disk galaxies as a function of their formation
redshift, as well as the location and luminosity of the ionizing sources
within the disks.

\subsection{Geometry}

We adopt the theoretical properties of disks forming within cold dark
matter halos (Mo et al. 1998; Navarro et al. 1997).  A dark matter halo of
mass $M_{\rm HALO}$ which forms at a redshift $z_{\rm f}$ is characterized
by a virial radius (Navarro et al. 1997),
\begin{equation}
r_{\rm vir}=0.76\left( {{M_{\rm HALO}} \over {10^8
h^{-1}M_\odot}}\right)^{1/3} \left( {{\Omega_0}\over {\Omega(z_{\rm f})}}
{{\Delta_c}\over {200}} \right)^{-1/3} \left({{1+z_{\rm f}}\over {10}}
\right) h^{-1}\, {\rm kpc}\; ,
\end{equation}
where $H_0 = 100 h$km~s$^{-1}$~Mpc$^{-1}$ is the Hubble constant,
$\Omega_0$ is the present mean density of matter in the universe in units
of the critical density ($\rho_{\rm crit} = 3H_0^2 / 8 \pi G$), and
\begin{equation}
\Omega(z_{\rm f}) = { {\Omega_0 (1+z_{\rm f})^3} \over
{\Omega_0 (1+z_{\rm f})^3 + \Omega_{\Lambda} +
(1-\Omega_0-\Omega_{\Lambda})(1+z_{\rm f})^2} }
\; .
\end{equation}
$\Delta_c$ is the threshold overdensity of the virialized dark matter
halo which can be fitted by (Bryan \& Norman 1998),
\begin{equation}
\Delta_c = 18\pi^2 +82d -39d^2\; ,
\end{equation}
for a flat universe with a cosmological constant, where $d=\Omega(z_f) - 1$.

We assume that the disk mass (stars plus gas) is a fraction $m_d$ of the
halo mass
\begin{equation}
M_{\rm DISK} = m_d M_{\rm HALO} \; ,
\end{equation}
where $m_d = \Omega_b/\Omega_0$, and $\Omega_b$ is the present baryonic
density parameter.  We adopt $\Omega_0 = 0.3$ and $\Omega_b = 0.05$, giving
$m_d = 0.17$.  At the high-redshifts of interest, most of the virialized
galactic gas is expected to cool rapidly and assemble into the disk.  Mo et
al. (1998) suggest values in the range $0.05<m_d<1$.  In our simulations in
\S~3.1 we also consider $m_d = 0.02$ which is an order of magnitude lower
than our canonical value of $m_d = 0.17$.  The exponential scale-radius of
the disk is given by (Mo et al. 1998)
\begin{equation}
R = \left({{j_d}\over{\sqrt{2} m_d}}\right)\lambda r_{\rm vir}\; ,
\end{equation}
where the disk angular momentum is a fraction $j_d$ of that of the halo,
and the spin parameter $\lambda$ is defined in terms of the total energy,
$E_{\rm HALO}$, and angular momentum, $J_{\rm HALO}$, of the halo, $\lambda
= J_{\rm HALO} |E_{\rm HALO}|^{1/2} G^{-1} M_{\rm HALO}^{-5/2}$.  For our
calculation of the escape of ionizing radiation we will adopt the values of
$j_d/m_d=1$ and $\lambda=0.05$, yielding $R= 0.035 r_{\rm vir}$. These
values provide a good fit to the observed size distribution of galactic
disks, given the characteristic value of the spin parameter found in
numerical simulations of halo formation (Mo et al. 1998, and references
therein).

The vertical, $z$, structure of thin galactic disks is dictated by their
self-gravity.  For simplicity, we assume that the disk is isothermal and
its surface density varies exponentially with radius. The number density of
protons in the disk is then given by (Spitzer 1942),
\begin{equation}
n(r,z) = n_0 {\rm e}^{-r/R} {\rm sech}^2 \left({ {z}\over {\sqrt{2}
z_0}}\right)\; .
\label{eq:6}
\end{equation}
where
\begin{equation}
z_0 = { {c_s} \over ({{4\pi G \mu m_H n_0 {\rm e}^{-r/R}}})^{1/2}}\; ,
\label{eq:z_0}
\end{equation}
is the scale height of the disk at at radius $r$, $\mu$ is the
atomic weight of the gas and $m_{\rm H}$ is the mass of a hydrogen atom.
Here $c_s={\sqrt{kT/\mu m_H}}$ is the sound speed (or the effective
turbulence speed), which dictates the scale-height of the disk.  We take
$c_s=10~$km~s$^{-1}$, which corresponds to a gas temperature of $\sim
10^4$~K. This temperature is typical of photoionized gas, and should
characterize the galactic disks of interest here since the atomic cooling
rate decreases strongly at lower temperatures. The combination of
isothermality and the exponential radial profile results in a disk
scale-height that increases with radius [see Eq.~(\ref{eq:z_0})]. The
galactic center number density, $n_0$, can be related to the total mass of
the disk which is obtained by integrating the density over the entire disk
volume
\begin{equation}
n_0 = { {M_{\rm DISK}} \over {\mu m_{\rm H} \int (n/n_0) 2\pi rdr
dz} } = { {G M_{\rm DISK}^2 } \over {128 \pi \mu m_{\rm H} c^2 R^4} }\; .
\label{eq:n_0}
\end{equation}
In Figure~1 we show the formation-redshift dependence of the number density
$n_0$, scale-radius $R$, and the ratio $z_0(r=R)/R$ for disks of different
masses.  To compare with the Milky Way, we examine the results for $M_{\rm
halo}=10^{12}M_\odot$ and $z_{\rm f}=1$.  This gives $n_0 \approx
100$cm$^{-3}$, a scale length $R\approx 5$kpc, and $z_0(r=R)/R \approx
0.002$, yielding a disk that is denser and thinner than that of the Milky
Way.  The discrepancy results from our choice of $m_d = 0.17$, whereas for
the Milky Way the disk mass today is a reduced fraction $m_d \sim 0.05$ of
the halo mass, possibly due to the effects of supernovae driven outflows or
inefficient cooling of the cosmic gas (the latter phenomenon could be
important at $z\la 2$ when the IGM becomes rarefied and hot, but is likely
to be irrelevant for galaxies which form at higher redshifts out of the
much denser and cooler IGM).  A value of $m_d = 0.05$ indeed yields
$n(r=2R)\approx 1$cm$^{-3}$ and $z_0(r=2R)/R \approx 0.04$, more
appropriate for the mean baryonic density in the Solar neighborhood.

We emphasize that fragmentation of the gas into stars or quasar black holes
is only possible as a result of substantial cooling of the gas well below
its initial virial temperature.  Since molecular hydrogen is likely to be
photo-dissociated in the low metallicity gas of primeval galaxies (Haiman,
Rees, \& Loeb 1997; Omukai \& Nishi 1999), only atomic cooling is
effective, and so stars and quasars are likely to form only in halos with a
virial temperature $\gg 10^4$ K, where atomic cooling is effective.  We
therefore restrict our attention to such halos. Inside such halos, thin
disks can exist for our assumed gas temperature of $\sim 10^4~{\rm K}$.

We restrict our simulations to halos in the range $10^9M_\odot$ to
$10^{12}M_\odot$. The gas in lower mass halos is either unable to cool and
form stars (due to the ease by which the only available coolant, ${\rm
H_2}$, is photo-dissociated), or is boiled out of the shallow gravitational
potential wells of the host halos by photo-ionization heating of hydrogen
to $\ga 10^4$K (above the virial temperature of these halos) as soon as a
small number of ionizing sources form (Omukai \& Nishi 1999; Nishi \& Susa
1999; Barkana \& Loeb 1999a).  If low-mass halos lose their gas quickly,
then their contribution to the ionizing background can be evaluated
trivially from their assumed star formation efficiency, with no need for a
detailed radiative transfer calculation.  The most popular cosmology at
present involves a $\Lambda$CDM power spectrum of density fluctuations with
$\Omega_m =0.3$, $\Omega_\Lambda =0.7$, $\sigma_8 = 0.9$, $h=0.7$, and
$n=1$ (scale invariant spectrum).  For this cosmology we find that halos
with a dark matter mass of $10^9M_\odot$ are $1.4\sigma$ fluctuations at
$z=5$ and $2.5\sigma$ fluctuations at $z=10$.  Halos of mass
$10^{12}M_\odot$ are $2.9\sigma$ fluctuations at $z=5$ and $5.3\sigma$
fluctuations at $z=10$.

\subsection{Illumination}

We consider two separate cases for the sources of ionizing radiation within
the galactic disks: stars and quasars.  Below we describe the luminosity
and the spatial location of the sources in these different cases.

\subsubsection{Stars}

In Monte Carlo simulations of the transfer of starlight through galaxies,
the stellar sources are often represented by a smooth spatial distribution
(e.g., Wood \& Jones 1997; Ferrara et al. 1999, 1996; Bianchi, Ferrara, \&
Giovanardi 1996) rather than by individual point sources (but see recent
work by Cole, Wood, \& Nordsieck 1999; Wood \& Reynolds 1999).  In our
simulations, we consider two smooth distributions for the stellar
emissivity: $j_{\star}\propto n(r,z)$ and $j_{\star}\propto n(r,z)^2$.  In
the first case, we are assuming that the star formation efficiency is
independent of density while in the second case the stars are assumed to
form preferentially in denser gaseous regions. Note that the second
prescription reproduces the observed Schmidt law for the star formation
rate of spiral galaxies as a function of the disk surface density
(Kennicutt 1998).

We assume a sudden burst of (metal poor) star formation with a Scalo (1986)
stellar mass function, and adopt a corresponding ionizing luminosity of
$Q_\star(H^0) = 10^{46} M_\star / M_\odot$~s$^{-1}$, where $M_\star$ is the
stellar mass of the galaxy (Haiman \& Loeb 1997).  This luminosity is
consistent with detailed models of starburst galaxies (e.g., Leitherer et
al. 1999).  The stellar mass is defined by the fraction of gas which gets
converted into stars, $M_\star = f_\star M_{\rm DISK}$. We consider two
cases (independent of formation redshift), namely $f_\star = 0.04$ and
$f_\star = 0.4$, with the gaseous disk being reduced in mass by the factor
$1-f_\star$.  The value $f_\star = 0.04$ was used by Haiman \& Loeb (1997)
so as to reproduce the observed metallicity of $\sim 0.01 Z_\odot$ in the
IGM at $z\sim 3$ (Tytler et al. 1995; Songaila \& Cowie 1996).  
We also consider a higher value of $f_\star=0.4$, in case $\sim 90\%$ of
all the metals are retained within their host galaxies and only $\sim 10\%$
are mixed with the IGM. Since the actual IGM metallicity at $z\sim 3$ might
be in the lower range of $0.1$--1\%$Z_\odot$ (Songaila 1997), our assumed
values for $f_\star$ and the corresponding ionizing fluxes might be
considered as high.  The corresponding ionization fraction of the disk is
overestimated, and our derived escape fractions should therefore be
regarded as upper limits.

In summary, the total ionizing luminosity for stellar sources in our
simulations is related to the halo mass via
\begin{equation}
Q_\star (H^0)=   10^{46} m_d f_\star { {M_{\rm HALO}}\over
{M_\odot}}\, {\rm s}^{-1} \; .
\label{eq:ion_lumi}
\end{equation}

Our code assumes a constant value for this ionizing luminosity and solves
for the ionization structure of the disk and the consequent escape fraction
in a steady state.  Since the characteristic time over which a starburst
would possess the ionizing luminosity in equation~(\ref{eq:ion_lumi}) is
$\sim 3 \times 10^{6}~{\rm yr}$ for a Scalo (1986) stellar mass function
(see Fig. 4 in Haiman \& Loeb 1997), the star formation rate required in
order to maintain a steady ionizing luminosity of this magnitude is,
\begin{equation}
\dot{M}_\star\approx {f_\star M_{\rm DISK}\over 3\times 10^6~{\rm yr}} =23
{M_\odot\over {\rm yr}}\left({f_\star \over 0.04}\right)\left({M_{\rm
HALO}\over 10^{10}~M_\odot}\right).
\label{eq:SFR}
\end{equation}
This is a rather high star formation rate, as it implies that for
$f_\star\sim 4\%$ the entire disk mass will be converted into stars in less
than a Hubble time at $z\sim 10$. The assumed star formation rates are
unreasonably high for $f_\star=40\%$ and large halo masses (e.g.,
$\dot{M}_\star\sim 2\times 10^4~{M_\odot~{\rm yr^{-1}}}$ for $M_{\rm
HALO}\sim 10^{12}M_\odot$). For this reason, one may regard our calculated
escape fractions as upper limits.

In order to perform the photoionization calculation with our Monte Carlo
code we need to specify the flux-weighted mean of the opacity of neutral
hydrogen for a given ionizing spectrum 
(see \S~2.3 and the Appendix).  We adopt a
flux mean cross-section of $\bar{\sigma } = 3.15 \times 10^{-18}$~cm$^2$, 
which is appropriate for the composite Scalo IMF 
emissivity spectrum presented in Fig.~3 of Haiman \& Loeb (1997).

\subsubsection{Quasars}

In the case of a quasar we assume that the ionizing radiation emanates from
a point source at the center of the disk.  Haiman \& Loeb (1998, 1999) have
demonstrated that the observed B-band and X-ray luminosity functions of
high-redshift quasars at $z\ga 2.2$ can be fitted by a $\Lambda$CDM model
in which each dark matter halo harbors a black hole with a mass
\begin{equation}
M_{\rm BH}=m_{bh} M_{\rm HALO} \; ,
\label{eq:11}
\end{equation}
which shines at the Eddington luminosity, $L_{\rm Edd} = 1.4 \times 10^{38}
(M_{\rm BH}/M_\odot)$~ergs~s$^{-1}$, for a period of $\sim 10^6$ years when
the halo forms.  The required value of $m_{bh} \approx 6 \times 10^{-4}$,
is consistent with the inferred black hole masses in local galaxies
(Magorrian 1998). We will adopt this prescription in deriving the quasar
luminosity within a halo of a given mass.  As there is a large scatter
around the mean in the observed distribution of the black hole to bulge
mass ratio, we also consider lower values in the range $6\times
10^{-7}<m_{bh}<6\times 10^{-4}$ for a $10^{10}M_\odot$ halo (see \S~3.2).

The ionizing component of quasar spectra is not well determined
empirically; we use the calibration of Laor \& Draine (1993) and relate the
ionizing luminosity to the total bolometric luminosity by $Q_{\rm QSO}(H^0)
= 6.6 \times 10^9 (L_{\rm Edd}/~{\rm erg~s^{-1}})$~s$^{-1}$.  Hence, the
ionizing luminosity of a quasar in our models is related to its host halo
mass through
\begin{equation}
Q_{\rm QSO}(H^0)= 9.24\times 10^{47}m_{bh} { {M_{\rm HALO}}\over {M_\odot}}\,
{\rm s}^{-1} \; .
\label{eq:12}
\end{equation}
We consistently adopt a flux-weighted mean of the opacity for the
characteristic quasar spectrum of $\bar{\sigma } = 2.05 \times
10^{-18}$~cm$^2$ (Laor \& Draine 1993).

\subsection{Radiation Transfer}

In our simulations we employ a three dimensional Monte Carlo radiation
transfer code (Wood \& Reynolds 1999) that has been modified to include
photoionization.  Our photoionization code is similar to that of Och et
al. (1998), but somewhat simplified in that we only treat the
photoionization of hydrogen at a constant temperature.  However, as our
simulations are run on a three dimensional grid we are able to investigate
complex geometries which are relevant to the present study.  We track the
propagation of ionizing photons from the source as they are absorbed and
reemitted as diffuse ionizing photons or as non-ionizing photons (in which
case they are subsequently ignored).  In each computational cell the
contributions of each ionizing photon to the mean intensity are tallied, so
that we may determine the ionization fraction throughout our grid.  We
iterate on this procedure until the ionization fractions converge.  We keep
track of the number of ionizing photons that exit our grid either directly
from the source or after multiple absorption and re-emission as diffuse
ionizing photons, thus determining the escape fraction of ionizing
radiation.  Note that we consider escaping ionizing photons to be either
direct photons from the source or diffuse photons.  As with most models of
the escape of ionizing photons, we have not considered the detailed effects
of the ionizing spectrum and work with a total ionizing luminosity (e.g.,
Miller \& Cox 1993; Dove \& Shull 1994; Dove et al. 1999; Razoumov \& Scott
1999).  Where our work differs from ``Str\"omgren volume'' analyses is that
we calculate the ionization fraction throughout our grid and track diffuse
(absorbed and re-emitted) ionizing photons also. A more detailed
description of our code and its comparison with other photoionization
calculations are presented in the Appendix.

Our code is time independent and does not consider any time evolution of
the ionizing spectrum in the calculation of the escape fraction.  Stellar
evolution models imply that the ionizing luminosity remains fairly constant
for $\la 10^7$~years (a typical main sequence lifetime of an O star) and
decreases subsequently (Haiman \& Loeb 1997; Leitherer et al. 1999).  In
recent modeling of the Milky Way, Dove et al. (1999) showed that including
the time dependence (and also the effects of dynamical supershells) results
in a decrease of the escape fractions compared to their static calculations
(Dove \& Shull 1994).  Therefore we will tend to overestimate the escape
fraction since we ignore the time decay of the ionizing luminosity, and our
results will provide upper limits for the escape of stellar ionizing
radiation from smooth galactic disks.  For quasars, Haiman \& Loeb (1998)
found that the observed quasar luminosity functions are best fit with a
quasar lifetime of $\sim 10^6$yr.  As for the stellar case, our static
quasar models yield upper limits on the escape fraction. However, since the
maximum ionizing luminosity of quasars is higher by $\ga 2$ orders of
magnitude than that of stars and involves harder photons, and since it all
originates from a single point, the propagation of the ionization front is
expected to be much faster for quasars, and so the steady-state assumption
should, in fact, be more adequate in their case.

For both stars and quasars, we ignore at first the effects of energetic
outflows on the galactic disk.  Outflows could clear away material and
raise the calculated escape fractions. We will incorporate this dilution
effect by artificially reducing the absorbing gas mass in the disk in some
of the numerical runs.

\section{Results}

\subsection{Stars}

We calculated the escape fraction of stellar ionizing radiation by
simulating the disk out to a radius of $3R(z_{\rm f})$, and using a smooth
emissivity profile in it.  As the disk density increases with redshift, the
recombination rate increases and the ionizing luminosity of the galaxy
results in a lower ionization fraction, leading to a decrease in the escape
fraction with increasing redshift.

Figure~2 shows the escape fraction as a function of formation-redshift for
various halo masses, assuming $f_\star=4\%$ and emissivities $\propto n^2$
(Fig.~2a), and $\propto n$ (Fig.~2b).  In these models the largest escape
fractions occur for low mass disks, but the escape fractions are $f_{\rm
esc} < 0.1$\% at $z_{\rm f}>5$.  When the emissivity scales as $\propto
n^2$, the sources are embedded more deeply within the disk, leaving a
larger column of hydrogen to be ionized and thus leading to smaller escape
fractions than in the case where the emissivity is $\propto n$.

The efficiency of converting baryons into stars is a free parameter,
$f_\star$, in our models.  In Figure~3 we show escape fractions for the
much larger efficiency, $f_\star = 40$\%.  As expected, the escape
fractions are larger than the corresponding values shown in Figure~2 due to
the increased ionizing luminosity [see Eq.~(\ref{eq:ion_lumi})] and the
reduced mass of the gaseous disk in the simulations.  However, the escape
fractions are very small at high redshifts, certainly much smaller than the
value of $f_{\rm esc} \approx 50$\% adopted by Madau (1999) as necessary
for stellar sources to keep the intergalactic medium ionized at $z\sim 5$.

In Figure~4 we show results for a simulation in which $f_\star = 4$\%, but
with an order-of-magnitude reduction in the density of the smooth gaseous
disk, $n_0 \rightarrow n_0/10$.  Since we keep the scale height of the
smooth gas unchanged, this is equivalent to lowering the mass of the
absorbing gas in the disk by a factor of ten, and may result from the
incorporation of gas into highly dense and compact clumps (see \S~4)
or expulsion of gas by stellar winds and supernovae.
The baryonic mass fraction in local disk galaxies, such as the Milky Way
galaxy, could have been influenced by outflows, as it is a few times lower
than the standard cosmic baryonic fraction of $m_d=0.17$ (see Fig.~5).  The
role of feedback from outflows is expected to be more pronounced at high
redshifts, where the characteristic potential well of galaxies is
shallower. Note, however, that under these conditions the overall star
formation efficiency is expected to be reduced.  Nevertheless, in our
calculation we left the stellar luminosity unchanged and only reduced the
mass of the smoothly distributed gas; under these circumstances,
$f_\star=4\%$ corresponds to the stars having $40\%$ of the smooth gaseous
disk mass.  This calculation was intended to simulate the most favorable
conditions for the escape of ionizing radiation from high-redshift
disks. Figure~4 illustrates that although the resulting escape fractions
for this extreme case are larger than the corresponding results in
Figure~2, they are still negligible for halo masses $M_{\rm halo}>
10^{10}M_\odot$ at $z_{\rm f} > 10$. It should be further noted that these
low escape fractions were calculated for the steady ionization state of the
disk implied by the rather high star formation rates in equation~(10).  We
also examined the case of gas removal from the disk due to outflows,
resulting in an increased disk scale height.  Escape fractions for reducing
the central density, $n_0$, by a factor of ten (or correspondingly the disk
mass by a factor of three in this case) yield results very similar to those
in Figure~4.

In order to investigate the effects on the escape fractions of a lower 
disk mass than our assumed $m_d=0.17$ we have performed a simulation for 
a $10^{10}M_\odot$ halo assuming $f_\star =4\%$, $j_\star\propto n$, but 
with $m_d=0.02$.  The results are shown in Fig.~5.  The lower disk mass 
yields a lower ionizing luminosity (Eq.~9) and a lower gas density (Eq.~8).  
The overall result of this is an increase in the escape fraction, albeit 
for galaxies of intrinsically lower luminosity.

\subsection{Quasars}

Being a point source of ionizing radiation, a quasar is expected to create
an ionized region around the center of the galactic disk.  The disk
geometry presents a higher opacity to the quasar's ionizing photons along
the disk midplane than it does perpendicular to it.  If the quasar
luminosity is sufficiently high, it creates a vertically extended
photoionized region of low opacity through which ionizing photons may
escape.  We have therefore restricted our simulations of quasar sources to
the innermost region of the disk and performed the radiation transfer
within a cube of size $40z_0(r=0)$ on a side.

Figure~6 presents the escape fraction as a function of formation-redshift
for various halo masses.  The escape fractions are typically in the range
of 20\%--80\% over a wide range of formation-redshifts and halo masses.
They decrease as the disks get denser with increasing formation-redshift,
but are always significantly larger than for the stellar case.  In all our
simulations, the vertical escape routes are created by the ionizing
luminosity of the quasar without considering any additional dynamical
clearing of the gas by outflows (e.g., a jet or a wind) from the quasar.
Such dynamical action could open up low density escape routes for the
ionizing photons, thereby increasing the already high escape fraction.  In
Figure~6b we show escape fractions assuming the gaseous disk has a density
smaller by a factor of ten than considered in Figure~6a ($n_0 \rightarrow
n_0/10$).  As expected, the escape fractions are very high in this case.

Our derivation of high escape fractions for quasars is consistent with the
lack of significant absorption beyond the Lyman limit at the host galaxy
redshift in observed quasar spectra (see, e.g., review by Koratkar \& Blaes
1999).

We have also simulated lower luminosity quasars ionizing a disk within a
$10^{10}M_\odot$ halo.  Figure~7 shows the escape fractions for black hole
masses in the range $6\times 10^{-7}<m_{bh}<6\times 10^{-4}$.  Since the
ionizing luminosity is proportional to the black hole mass [see
Eqs.~(\ref{eq:11}) and (\ref{eq:12})], the escape fraction decreases for
low luminosity quasars.  For very low luminosity quasars $f_{\rm esc}$
diminishes because the Str\"omgren sphere created by the quasar is smaller
than the scaleheight of the disk and all ionizing photons are trapped. In
these cases, the emergence of hydrodynamic outflows or jets from the quasar
accretion flow could be important in opening escape channels for the
ionizing radiation.

\subsection{Angular Distribution of Escaping Ionizing Flux}

In addition to the ionization structure and escape fraction, our code also
provides the angular distribution of the escaping ionizing flux.  In
Figure~8 we show this angular distribution for stellar sources ($f_\star =
4$\%) or a central quasar within a halo of mass $M_{\rm halo} =
10^{10}M_\odot$ at $z_f = 8$.  The escape fraction for the stellar case is
$f_{\rm esc} = 6$\% and for the quasar $f_{\rm esc} = 60$\%.  Since the
system is axisymmetric in both cases, we show the normalized flux of
ionizing photons as a function of $\cos i$, where $i$ is the disk
inclination angle. The flux is normalized to unity for a face-on viewing of
the disk.  The figure clearly illustrates the asymmetry of the emerging
ionizing radiation field due to the dense disk, as photons escape
preferrentially along the paths of lower optical depth.  For the quasar
case the pole-to-equator flux ratio is more extreme since the central
source opens up a low opacity ionized chimney and is unable to ionize the
disk to a large distance in the midplane.  For stars, the emission is
distributed throughout the disk resulting in a more moderate
pole-to-equator flux variation.  The cosmological H~II regions formed by
both types of sources in the surrounding IGM will show neutral shadow
regions aligned with the disk midplane.

\section{Effects of Clumping on the Escape of Ionizing Radiation from 
Stars}

In \S~3.1 we calculated the escape fraction, $f_{\rm esc}$, for stellar
sources, assuming that the ionizing sources and absorbing galactic gas were
smoothly distributed according to the formulae presented in \S~2.2.  In
this section we present results of simulations where either the gas, the
sources or both, have a clumpy distribution within the galactic disks.
Several recent papers have studied the dust scattering of radiation in a
two-phase medium, with emphasis on the penetration and escape of
non-ionizing stellar radiation from clumpy environments (Boisse 1990; Witt
\& Gordon 1996, 2000; Bianchi et al. 2000; Haiman \& Spaans 2000).  Dove et
al. (2000) investigated the escape of ionizing photons from the Milky Way
disk where inhomogeneities in the interstellar medium were modeled as
either spheres or cylindrical disks.  In all of these studies, clumping
allowed photons to penetrate to greater depths than in a smooth medium and
the escape of radiation was enhanced relative to the case where the same
gas mass was distributed smoothly.

In evaluating the escape of ionizing radiation from clumpy environments we
adopt the two-phase prescription from Witt \& Gordon (1996).  The two-phase
medium has two parameters, namely the volume filling factor of dense
clumps, $ff$, and the density contrast between the clump and interclump
medium, $C$.  We transform our smooth density distribution
[Eqs.~(6), (\ref{eq:z_0}), and (\ref{eq:n_0})] into a clumpy one
by looping through our three-dimensional grid and applying the following
algorithm in each grid cell
\begin{equation}
n_{\rm clumpy}=\cases 
{n_{\rm smooth}/[ff+(1-ff)/C], & if $\xi < ff$;\cr
n_{\rm smooth}/[ff(C-1)+1],&otherwise.\cr}
\end{equation}
where $\xi$ is a uniform random deviate in the range (0,1).  This algorithm
assures that {\it on average} the total disk mass is the same for the
clumpy and smooth models, and that the ensemble average of the clumpy
distribution follows the same spatial profile as the smooth gas does.  In
this approach the smallest clump is a single cell in our density grid.  Our
resolution is set by the size of our density grid ($100^3$ cells), so that
each cell is a cube of size $3R/50$, where $R$ is the radial scalelength of
the disk. For simplicity, we consider a two-phase medium for which the
interclump medium has a zero density. Such a medium is described by a
single parameter only, $ff$. In our simulations we approximate this extreme
case by adopting a very large density contrast $C=10^6$.  Slices through
the resulting density distribution for various clump filling factors are
shown in Figure~9.  We have also applied the above algorithm to generate a
clumpy emissivity distribution, which could result, for example, from
clusters of young stars.

In Figure~10 we show results of clumpy simulations for the case of a
$10^{10}M_\odot$ halo at $z_f=10$, where $m_d=0.17$, $f_\star=4\%$, and
$j_\star\propto n$.  In our smooth simulations this yields $f_{\rm esc} <
10^{-3}$.  The figure shows the results of three different clumpy
simulations.  The first simulation assumes a smooth distribution for the
gas and a clumpy distribution for the emissivity, while the second examines
a clumpy gas and a smooth emissivity distributions.  In the third
simulation both the emissivity and the gas density are clumped, but they
are not spatially correlated (i.e., we generated the clumpy density and
emissivity with the above algorithm, but from different random number
sequences).

For the first simulation, $ff$ is defined to be the filling factor of the
clumpy emissivity. We find that $f_{\rm esc} > 1\%$ is attained for $ff <
0.05$.  For these very low values of $ff$, the emissivity is concentrated
into less than 5\% of the galactic volume.  Within these compact
concentrations there is a higher flux of ionizing photons per hydrogen atom
compared to the smooth simulations.  Each high emissivity cell can now
generate a bigger Str\"omgren volume and, depending on its location within
the galactic disk, can open up an HII escape channel and yield a larger
$f_{\rm esc}$ than the smooth simulation does.  As the emissivity filling
factor approaches unity, we retrieve the very small $f_{\rm esc}$
characteristic of the smooth case.

The escape fractions are larger when the interstellar medium is clumped.
We find that with the above prescription for the random clumping the
results are insensitive to whether the emissivity is smooth or clumped.
(However, when the clumpy emissivity and density are correlated, i.e.,
emission arises only within the dense clumps, we find that the escape
fraction is negligible since the emission within the dense clumps is unable
to ionize the clumps and escape.)  When the clump filling factor is small,
there are lines of sight from either the smooth or clumpy emissivity that
do not encounter dense gas and the ionizing photons are free to directly
exit the galactic disk.  The escape fraction approaches unity for very
small clump filling factors.  Such small filling factors represent the
unphysical situations of all the galactic mass residing in a few dense and
highly compact clumps.  However, we do find $f_{\rm esc}$ can exceed
several percent for $ff\approx 20\%$ which is comparable to the filling
factors of H~I (Bregman, Kelson, \& Ashe 1993) and H~II (Reynolds et
al. 1995) in the Milky Way galaxy.  While the two-phase density
distribution we have adopted does not fully describe the hierarchical
structures observed in galactic disks, it does provide an estimate of the
porosity levels and clump filling factors required to allow significant
escape fractions from high redshift galaxies.

\section{Discussion}

We have calculated the escape fraction of ionizing photons as a function of
formation redshift, $z_{\rm f}$, for disk galaxies of various masses.  In
our canonical model, the disk mass within a given dark matter halo was
assumed to be the cosmic baryonic mass fraction, since the initial cooling
of the virialized gas with a temperature $T\ga 10^4$~K, is expected to be
very rapid compared to a Hubble time at the high redshifts of interest (see
relevant cooling rates in Binney \& Tremaine 1987).  We find that for
smooth disks the escape fraction for bright quasars is in excess of $\sim
30$\% for most galaxies at $z_{\rm f}>10$, whereas stellar sources are
unable to ionize their host galaxies, yielding much smaller escape
fractions.  In our smooth disk models, stellar escape fractions in excess
of a percent are never achieved beyond $z_{\rm f}=10$ (see Figure 2) unless
the conversion efficiency of baryons into stars, $f_\star$, is large
(Figure 3), requiring unreasonably high star formation rates [see
Eq.~(\ref{eq:SFR})].  Even for the most extreme cases, we find that disks
which form within rarer dark matter halos more massive than
$10^{10}M_\odot$, have negligible escape fractions at $z_{\rm f} \ga 10$.

Clumping or expulsion of gas from the disks would tend to leave a diluted
interstellar medium, and allow for larger escape fractions.  We have
investigated the effect of gas expulsion by considering a smooth gaseous
disk in which the density is an order of magnitude lower than implied by
the cosmic baryonic fraction (see Figs. 4 and 5).  In this case we find
that the escape fractions are considerably increased, but again only disks
within low mass halos ($M_{\rm halo}\la 10^{10}M_\odot$) exhibit escape
fractions in excess of several percent at $z_{\rm f}\ga 10$.  Note that
this case requires an extreme star formation rate even with $f_\star=4\%$,
since $40\%$ of the smooth gas mass is already incorporated into stars.  If
the depletion of the smooth gas is due to outflows, then these flows would
be effective at increasing the escape fraction only if they clear out the
absorbing gas from the disk in less than a few million years, the
characteristic timescale over which the ionizing radiation is emitted by
massive stars. In this case, the overall star formation efficiency is
expected to be lower than we assumed due to the rapid depletion of the
galactic gas reservoir (and so the total contribution of the host galaxy to
the ionizing background would be low).  We have also investigated the
escape fractions from clumpy disks with a two-phase density distribution
and find that escape fractions larger than a few percent can be achieved if
the inter-clump regions (which were assumed to be devoid of absorbing
material) occupy $\ga 80\%$ of the disk volume.

In all our simulations we have calculated escape fractions for the
asymptotic steady ionization state of the disk.  Unless the star formation
rates maintain the high values implied by equation~(\ref{eq:SFR}) over
several generations of massive stars, our simulations overestimate the
escape fractions since they do not include the lower ionization state of
the surrounding gas due to the lower ionizing luminosity at both early and
late times.  Considering the Milky Way galaxy, Dove et al. (1999) have
shown that including the effects of a finite source lifetime and radiation
transfer through supershells created by stellar winds and superovae will
{\it decrease} the escape fractions compared to the smooth, static cases we
have considered.  We expect this effect to be less important for quasars
than it is for stars; due to their much higher brightness, quasars reach
steady ionization conditions on a time scale much shorter than their
expected lifetime (which is $\ga 10^6$ yr).

Our discussion focused on isothermal ($T=10^4$K) disks that form within
dark matter halos with masses, $M_{\rm halo} \ge 10^9 M_\odot$.  Lower mass
halos yield disk scale heights that are not smaller than their scale radii
(see Fig.~1), since their virial temperature is lower than our assumed gas
temperature of $10^4$K.  The virialized gas in such halos is unable to cool
via atomic transitions, and might fragment into stars only if molecular
hydrogen forms efficiently in it (Haiman et al. 1999).  However, the star
formation process in such systems is expected to be self-destructive and to
limit $f_\star$ to low values. If only a small fraction of the gas converts
into massive stars then molecular hydrogen would be photo-dissociated
(Omukai \& Nishi 1999). More importantly, any substantial photo-ionization
of the gas in these systems (which is required in order to allow for a
considerable escape fraction) would heat the gas to a temperature $\ga
10^4$ K and boil it out of the gravitational potential well of the host
galaxy, thus suppressing additional star formation. Winds from a small
number of stars or from a single supernova could produce the same effect.
It is, of course, possible that by the time the gas is removed from these
systems, they would have already formed stars that contribute to the
ionizing background. However, the star formation efficiency under these
circumstances is expected to be significantly lower than in higher-mass
galaxies.

Our derived escape fractions should be regarded as upper limits since more
absorption is likely to be added by neutral gas in the vicinity of the
galaxies, due to infalling material, gas in galaxy groups, or gas in nearby
intergalactic sheets and filaments. The surrounding gas density is expected
to decline rapidly with distance away from a galactic source (roughly as
$\propto 1/r^{2.2}$ in self-similar infall models [Bertschinger 1985]) and
so have its strongest absorption effect close to the source.  Large scale
numerical simulations of reionization (e.g., Gnedin 1999) cannot resolve
the small scales of interest due to dynamic range limitations, and are
forced to adopt ad-hoc prescriptions for the escape fraction from galaxies.
A natural extension of the radiation transfer calculation presented in this
work would be to embed a galactic source inside its likely intergalactic
environment based on high-resolution small-scale hydrodynamic simulations,
and to find the fraction of ionizing photons which would escape into the
more distant intergalactic space.

Our calculations can be tested by spectroscopic observations of high
redshift galaxies, such as the Lyman-break population at $z\sim 3$--5
(Steidel et al. 1999).  While a direct detection of flux beyond the Lyman
break might prove difficult, infrared observations of the H$\alpha$
emission from the halos of these galaxies might be used to constrain the
escape fraction of their ionizing luminosities.

The major implication of this work is that ionizing radiation from stars in
high-redshift disk galaxies is expected to be trapped by their surrounding
high-density interstellar medium unless ouflows or fragmentation dilutes
this gas by more than an order of magnitude.  If most of the gas remains in
the smooth disk during the lifetime of the massive stars ($\la 10^7~{\rm
yr}$), then stellar sources are unable to contribute significantly to the
intergalactic ionizing background at redshifts $z\ga 6$, and only
mini-quasars are capable of reionizing the Universe then.  If mini-quasars
are not sufficiently abundant or have very low luminosities at these
redshifts (see Fig. 7), then reionization must have occurred late, close to
the horizon of current observations at $z\sim 6$--7.

\acknowledgements

KW acknowledges support from NASA's Long Term Space Astrophysics Research
Program (NAG 5-6039) and thanks John Mathis, Jon Bjorkman, George Rybicki,
and Barbara Whitney for discussions related to photoionization within the
Monte Carlo radiation transfer code.  AL was supported in part by NASA
grants NAG 5-7039 and NAG 5-7768.  We thank Rennan Barkana, Benedetta
Ciardi, Andrea Ferrara, Marco Spaans, and Linda Sparke for discussions and
suggestions on this work.

\section*{Appendix --- Monte Carlo Photoionization}

Our three-dimensional Monte Carlo radiation transfer code (Wood \& Reynolds
1999) has been modified to include photoionization balance in a scheme
close to that of Och et al. (1998).  In this Appendix we describe the
simplifications that we have adopted and show comparisons of our code with
other calculations.

We assume that all ionizing photons encounter an average opacity (cross
section) on their random walk through our grid.  The opacity in each cell
is given by $n_{H^0} \bar{\sigma}$, where $n_{H^0}$ is the number density
of neutral hydrogen in the cell and $\bar{\sigma}$ is the average
cross-section.  The optical depth along a path length $l$ through the cell
is then $\tau = n_{H^0} \bar{\sigma} l$.  We use two different
cross-sections depending on whether the photons are directly emitted by the
source or have been absorbed and re-emitted as diffuse ionizing photons.
The cross-section we use for source photons is averaged over the flux,
$\bar{\sigma} = \int_{\nu_0}^\infty F\sigma d\nu / \int F d\nu$, where $F$
is the source ionizing spectrum, $\nu_0$ is the frequency at the Lyman
edge, and $\sigma$ is the absorption cross-section of hydrogen.  The
diffuse ionizing spectrum is strongly peaked at energies just above 13.6 eV,
and so we set the cross-section for the random walks of diffuse photons to
be the hydrogen cross-section at energies just above 13.6 eV, $\bar{\sigma}
= 6.2 \times 10^{-18}$cm$^2$.  This is the main simplification we have made
over the work of Och et al. (1998): we are essentially considering only two
frequencies in the radiation transfer.

The simulation starts with an assumed ionization fraction throughout the
grid and then source photons are emitted isotropically and the
cross-section is set to its flux averaged value.  Random optical depths are
chosen from $\tau = -\ln {\xi}$, where $\xi$ is a random number in the
range $0< \xi< 1$.  The photons are tracked along their propagation
direction until they reach an interaction point at a distance $s$ from
their emission (or re-emission) point, where $s$ is determined from $\tau =
\int_0^s n_{H^0} \bar{\sigma} dl$.  At the interaction points, the photons
are absorbed and re-emitted as either line photons with energies $h\nu <
13.6$ eV, or as diffuse ionizing photons with $h\nu > 13.6$ eV.  The
probability of a photon being re-emitted as an ionizing photon is the ratio
of the energy in the diffuse ionizing spectrum to the total energy [see
Eq.~(16) and the accompanying discussion in Och et al. 1998].  In our
simulations we use the following simplified form for this probability,
\begin{equation}
P= { {\alpha_1} \over {\alpha_A} } = 1- { {\alpha_B} \over {\alpha_A} }
\end{equation}
where $\alpha_1$ is the recombination coefficient to the ground state of
hydrogen (which gives the diffuse ionizing spectrum), $\alpha_B$ is the
recombination coefficient to the excited levels, and $\alpha_A$ is the
recombination coefficient to all levels, including the ground state.  In
our simulations we adopt recombination coefficients appropriate for a gas
at $10^4$K (Osterbrock 1989): $\alpha_A = 4.18 \times
10^{-13}$cm$^3$~s$^{-1}$, $\alpha_B = 2.59 \times 10^{-13}$cm$^3$~s$^{-1}$,
and thus $P=0.38$.  An absorbed photon is re-emitted isotropically from its
absorption point as a diffuse ionizing photon if $\xi < P$.  If $\xi > P$,
then the photon will be re-emitted as a non-ionizing photon, in which case
we ignore the opacity it encounters and remove it from the simulation.

As the photons are tracked on their random walks, we calculate their
contribution to the mean intensity in each cell, using the path length
formula from Lucy [1999, his Eq.~(13)],
\begin{equation}
J = { {E} \over {4\pi \Delta t V} }\sum l \; ,
\end{equation}
where the total ionizing luminosity is split up into $N$ equal energy
packets (Monte Carlo ``photons'') of luminosity $E/\Delta t=h\nu Q(H^0)/N$,
$l$ is the path length the photon traverses in each cell, and $V$ is the
volume of a cell in our grid.  In the photoionization equilibrium equation
\begin{equation}
n_{H^0}\int_{\nu_0}^\infty {{4\pi J_\nu}\over{h\nu}} \sigma_\nu d\nu =
\alpha_A n_e n_p \; ,
\end{equation}
we need the integral of ${{4\pi J_\nu\sigma_\nu }/{h\nu}}$ over all
ionizing frequencies. (Here, $n_e$ and $n_p$ are the number densities of
free electrons and protons.)  As we are effectively only considering the
cross-section at two frequencies, a counter is kept for each cell that
approximates the integral as
\begin{equation}
\int_{\nu_0}^\infty {{4\pi J_\nu}\over{h\nu}} \sigma_\nu d\nu =
{ {Q(H^0)}\over {N}} {{\sum l}\over {V}} \bar{\sigma}\; ,
\end{equation}
where $\bar{\sigma}$ is the mean cross section seen by either the source
photons or the diffuse photons.

All photons are tracked and their contributions to the mean intensity
tallied in each cell until they exit the grid either through multiple
absorption and re-emission as diffuse ionizing photons, or as non-ionizing
photons.  At this point we have the mean intensity within each grid cell
and can solve the hydrogen photoionization equation to determine the
ionization fraction in each cell.  With the ionization fraction and
therefore the opacity calculated throughout the grid we iterate the above
procedure, calculating new ionization fractions throughout the grid until
convergence is achieved.  We typically find that the ionization fractions
converge within five iterations.

In order to test the validity of our assumptions we have tested our code
against the spherically symmetric photoionization code of Mathis (1985).
The test case we have chosen is the same as that presented by Och et al.
(1998).  A blackbody ionizing source of $T_{\rm eff} = 4\times 10^4$K and
ionizing luminosity $Q(H^0) = 4.26\times 10^{49}$s$^{-1}$ ionizes a region
of constant number density $n_H = 100$cm$^{-3}$.  In our approach, the flux
mean cross section for this case is $\bar{\sigma}= 3\times 10^{-18}$cm$^2$.
Figure~11 shows the ionization fraction as calculated by our code and by
the Mathis code.  Even though we have employed many simplifications
compared to Mathis' code, our results for hydrogen photoionization
(ionization fraction and extent of the ionized region) are in excellent
agreement with his.

As a second test case we have compared our code with the two-dimensional
``Str\"omgren volume'' calculations of Dove \& Shull (1994, their Figs.~1 and
2).  They calculated the escape fraction and ionized volumes formed within
a three component galactic disk by point sources of different luminosity.
Figure~12 shows the escape fractions calculated by our Monte Carlo code and
the analytic escape fractions of Dove \& Shull [1994, their Eq.~(16)].  We
have plotted the total escape fraction from both sides of the disk, whereas
Dove \& Shull (1994) showed the escape fraction from one side of the disk.
Our Monte Carlo results are in excellent agreement at high luminosities,
but are systematically lower at low luminosities.  This discrepancy arises
because we calculate the ionization fraction throughout our grid, whereas
the Str\"omgren volume analysis assumes that the medium is either neutral or
fully ionized.  In Figure~13 we show three slices through our density grid
showing the total number density, ionization fraction, and locations of the
regions where photons are absorbed and re-emitted as escaping non-ionizing
radiation.  While the Str\"omgren volume is similar to that shown in
Figure~1d of Dove \& Shull (1994), our calculated neutral fraction of
$f\approx 10^{-3}$ provides sufficient opacity to absorb source photons,
thereby reducing the escape fraction compared to that of Dove \& Shull
(1994).  In the analysis of Dove \& Shull, source photons can escape
directly through this ionized ``chimney'' where they assume a vanishing
neutral fraction.

\begin{figure}[t]
\centerline{\plotfiddle{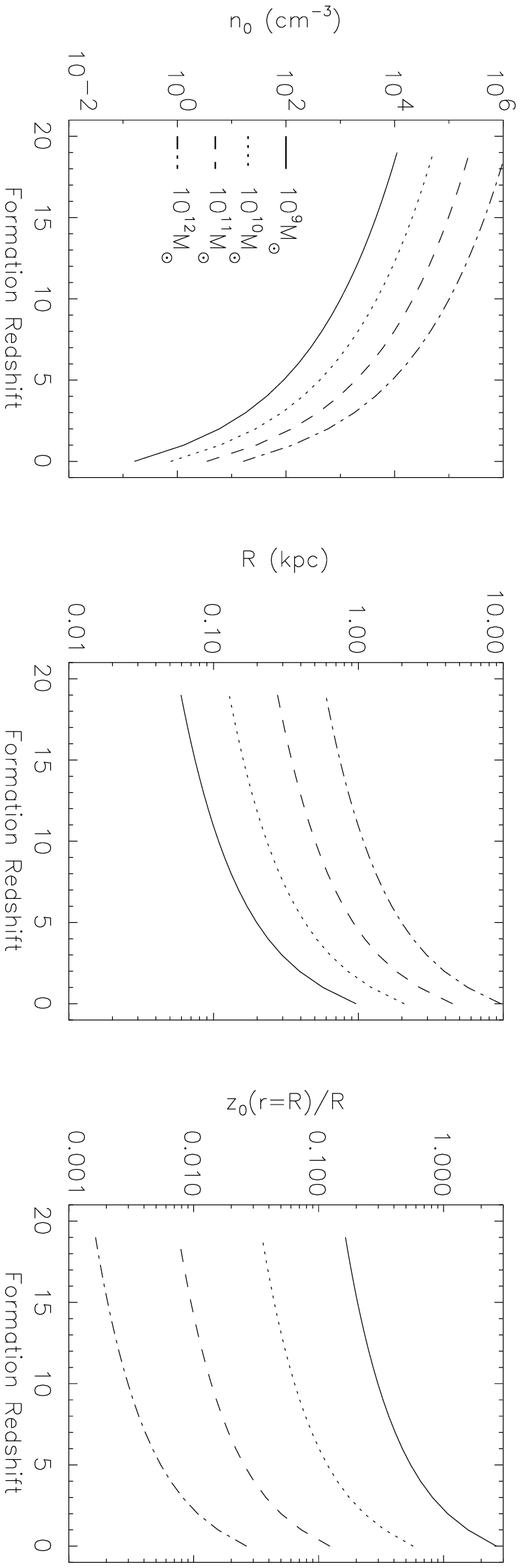}{6in}{90}{65}{65}{20}{100}}
\caption{Central densities, scale lengths, and the ratio of
scale-height/scale-length as functions of redshift, for our model of
isothermal, self gravitating disks that form within dark matter halos of
various masses.  }
\label{fig:1}
\end{figure}

\begin{figure}[t]
\centerline{\plotfiddle{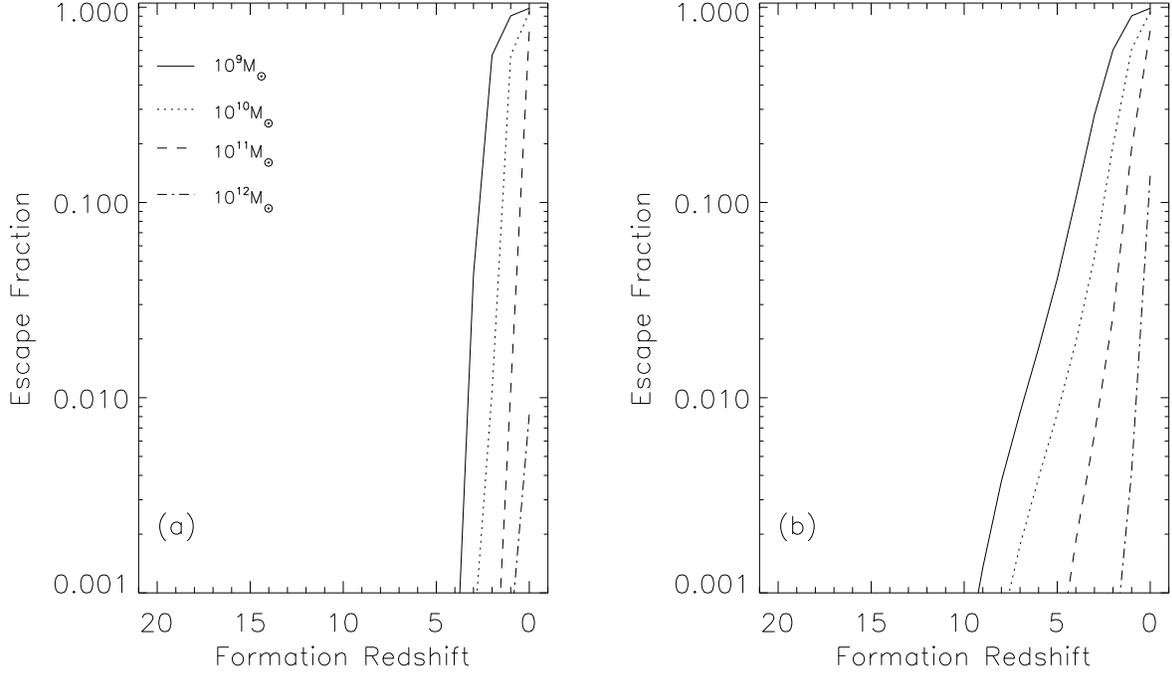}{6in}{90}{65}{65}{20}{100}}
\caption{Stellar escape fractions for smooth density disks within dark
matter halos of different masses, assuming $f_\star = 4$\%.  (a)~Ionizing
emissivity distributed $\propto n^2$.  (b)~Ionizing emissivity distributed
$\propto n$.}
\label{fig:2}
\end{figure}

\begin{figure}[t]
\centerline{\plotfiddle{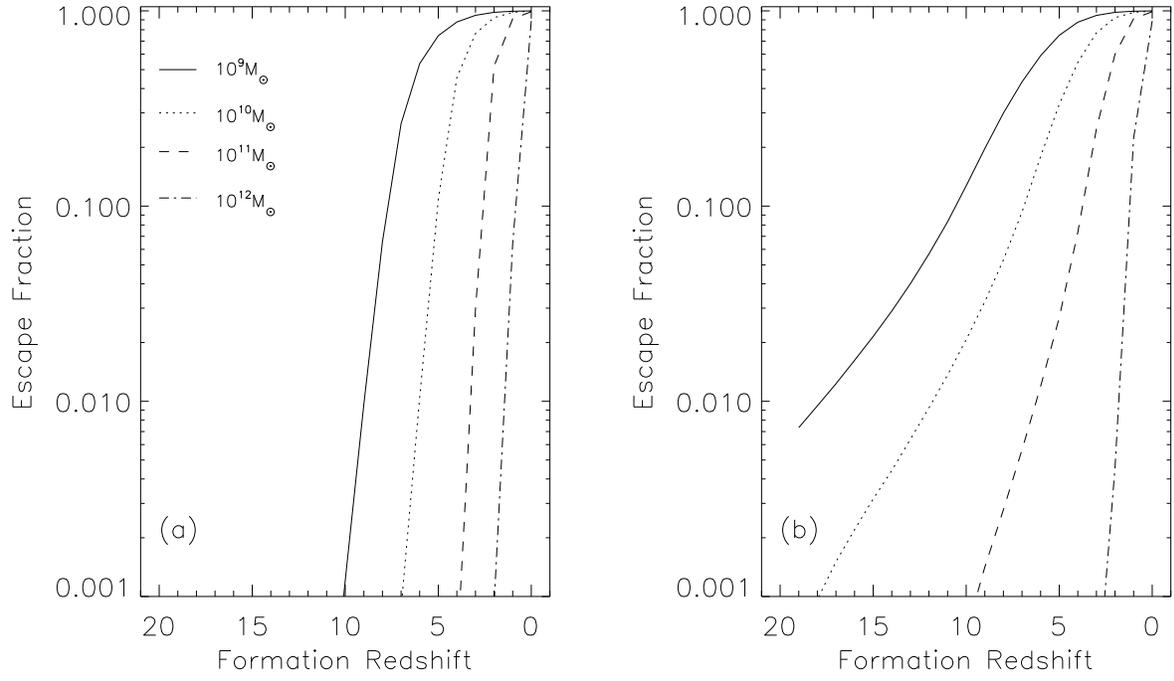}{6in}{90}{65}{65}{20}{100}}
\caption{Same as in Figure 2, but for
$f_\star = 40$\%.  }
\label{fig:3}
\end{figure}

\begin{figure}[t]
\centerline{\plotfiddle{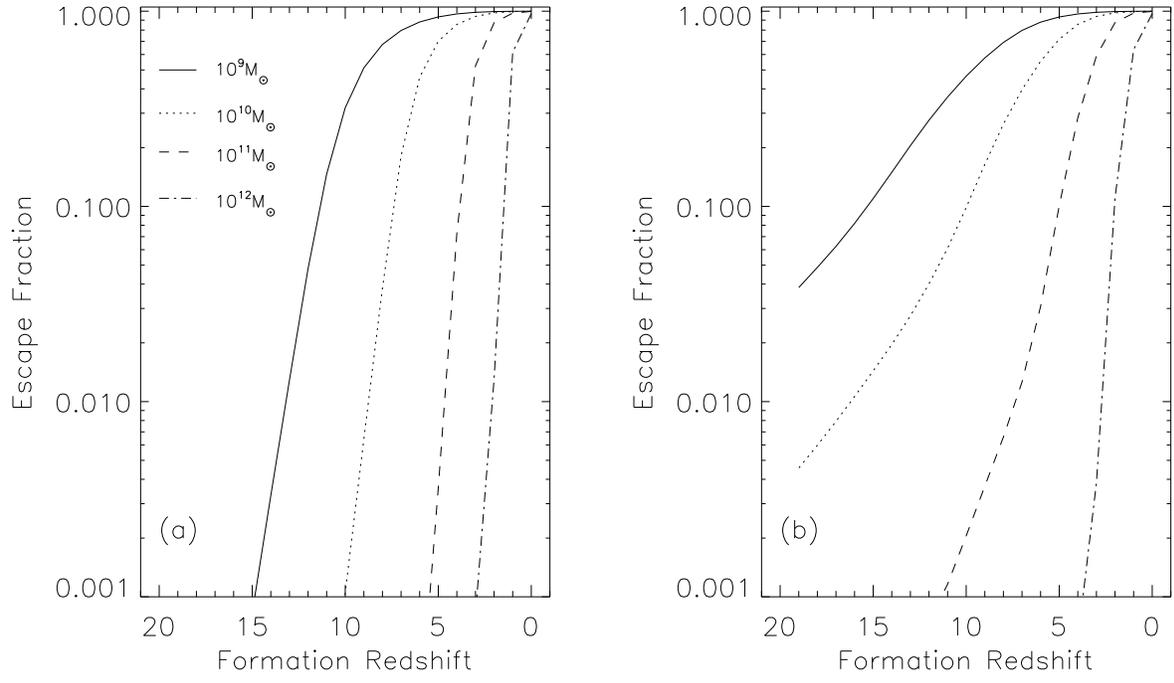}{6in}{90}{65}{65}{20}{100}}
\caption{Same as in Figure 2, but with an order of magnitude reduction in
the density of the smooth gaseous disk, or $n_0 \rightarrow n_0/10$.}
\label{fig:4}
\end{figure}

\begin{figure}[t]
\centerline{\plotfiddle{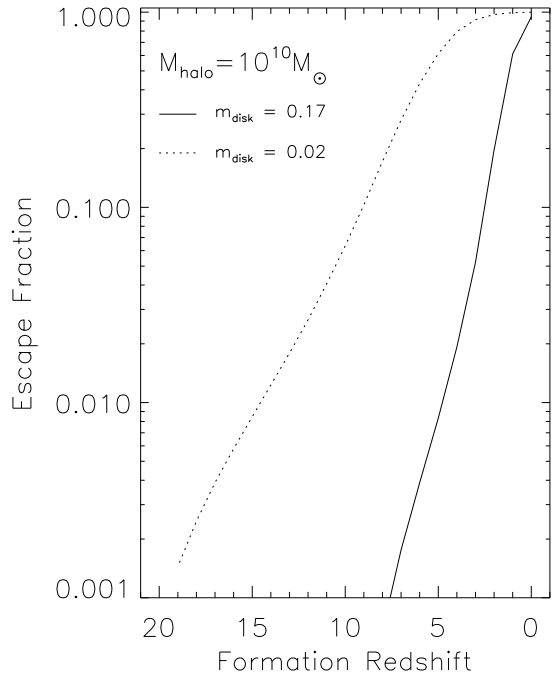}{6in}{90}{65}{65}{20}{100}}
\caption{Escape fractions of stellar ionizing photons from a smooth 
disk within a $10^{10}M_\odot$ halo.  The curves represent the cases 
when the ratio of disk mass to halo mass is $m_d=0.17$ and $m_d=0.02$.}
\label{fig:5}
\end{figure}

\begin{figure}[t]
\centerline{\plotfiddle{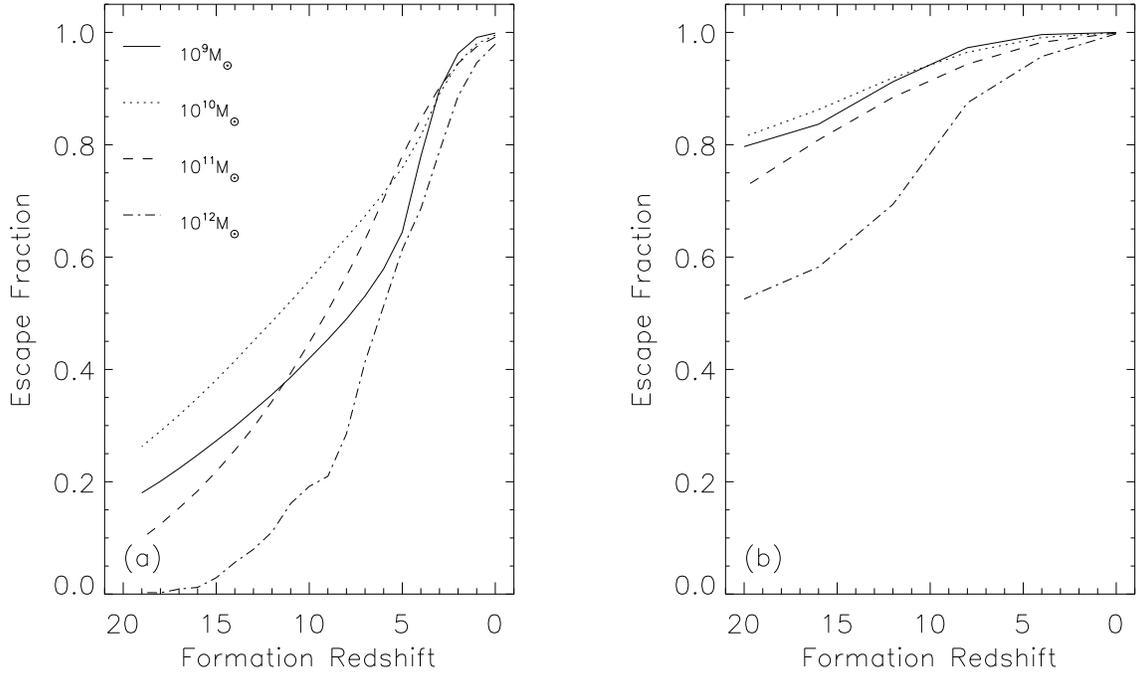}{6in}{90}{65}{65}{20}{100}}
\caption{Escape fractions for ionization of galactic disks within dark
matter halos of different masses by a central quasar: (a)~Disk number
density given by equation~(\ref{eq:n_0}); (b)~Diluted disk mass, $n_0
\rightarrow n_0/10$.}
\label{fig:6}
\end{figure}

\begin{figure}[t]
\centerline{\plotfiddle{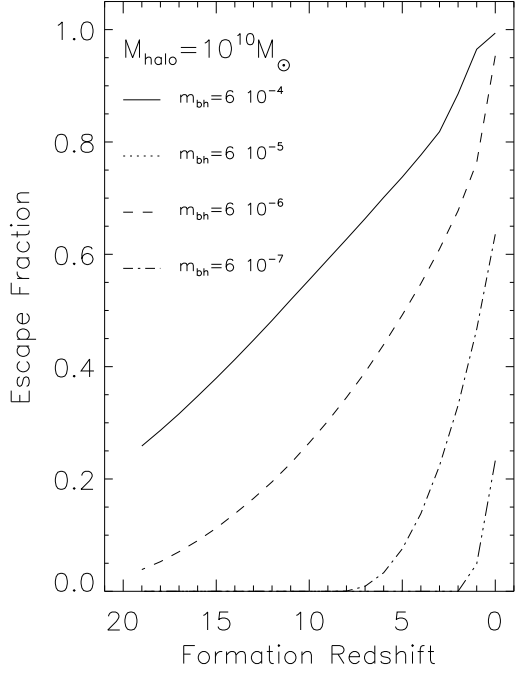}{6in}{90}{65}{65}{20}{100}}
\caption{Escape fractions of quasar ionizing photons from a smooth 
disk within a $10^{10}M_\odot$ halo.  The curves represent the cases 
when the ratio of black hole mass to halo mass is in the range 
$6\times 10^{-7}<m_{bh}<6\times 10^{-4}$.  The quasar luminosity is 
assumed to be proportional to the black hole mass, so the escape 
fractions are smaller for lower mass black holes.}
\label{fig:7}
\end{figure}

\begin{figure}[t]
\centerline{\epsfysize=5.7in\epsffile{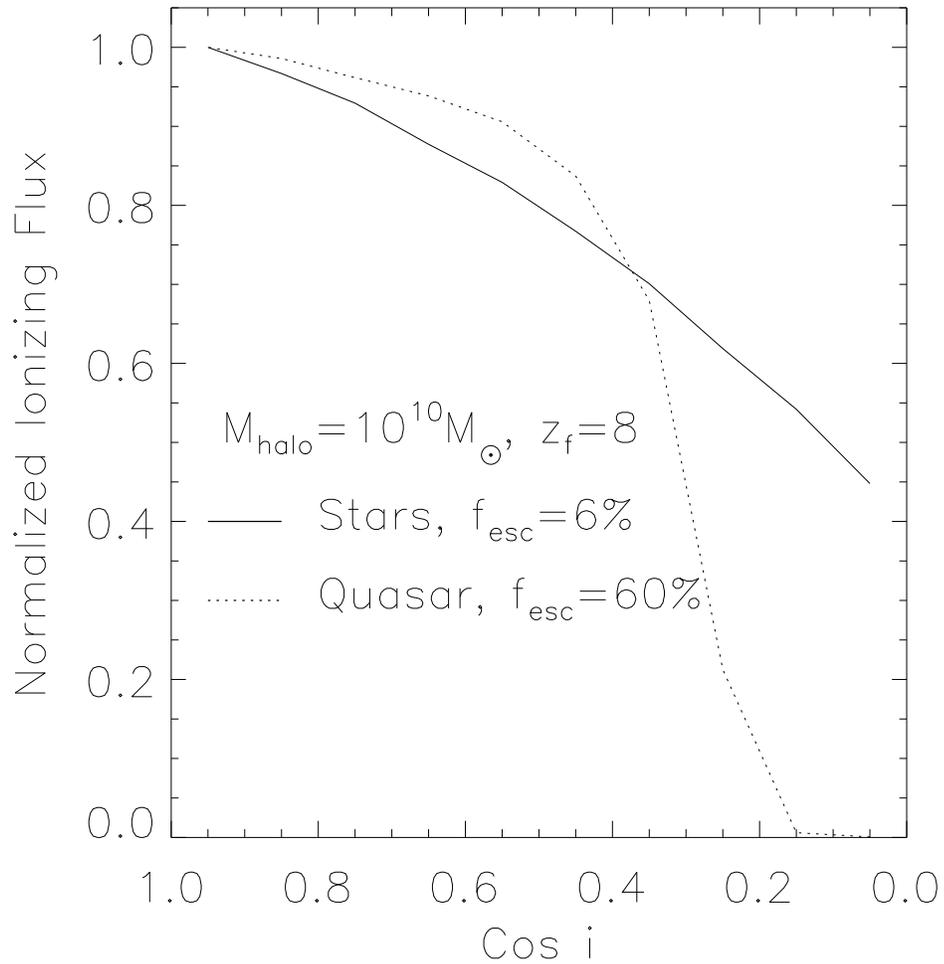 }}
\caption{Angular distribution of the escaping ionizing flux from a quasar
or stars within a dark matter halo of $M_{\rm halo} = 10^{10}M_\odot$ at
$z_f = 8$. The flux is arbitrarily normalized to unity for a disk
inclination angle $i=0^{\circ}$.}
\label{fig:8}
\end{figure}

\begin{figure}[t]
\centerline{\plotfiddle{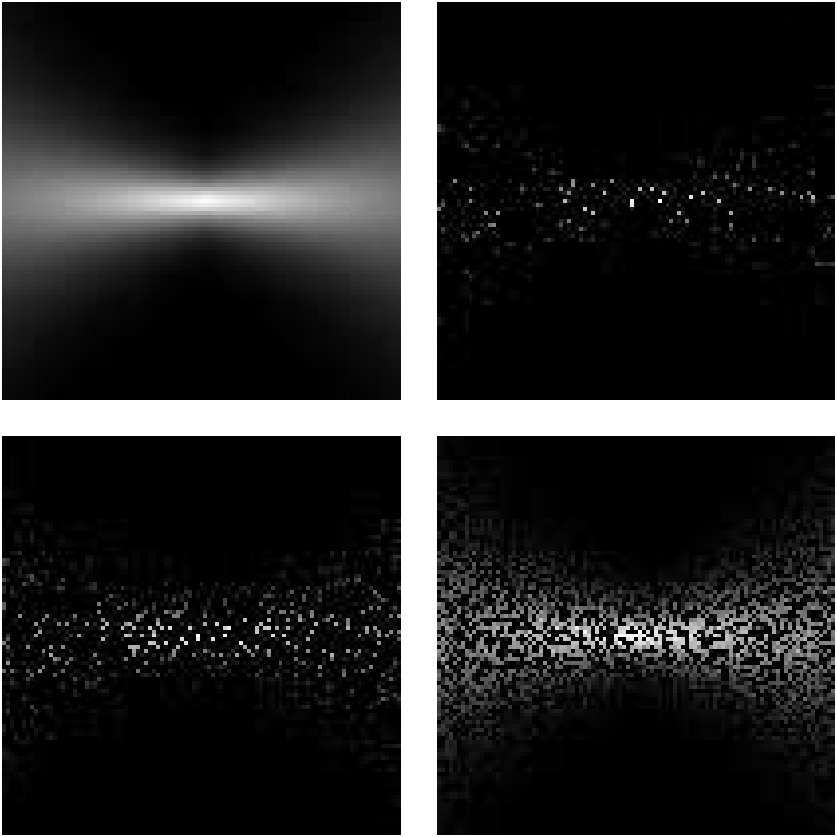}{6in}{0}{65}{65}{-450}{-50}}
\caption{Slices through a disk formed within a $10^{10}M_\odot$ halo at 
$z_f=10$.  The panels show the smooth case (upper left) and three clumpy 
cases.  The different clumpy models are for various volume filling factors 
of dense clumps.  Upper right panel: $ff=3\%$, lower left panel: $ff=10\%$, 
lower right panel: $ff=50\%$.  All panels show the fourth root of the 
number density $n^{1/4}$.}
\label{fig:10}
\end{figure}

\begin{figure}[t]
\centerline{\plotfiddle{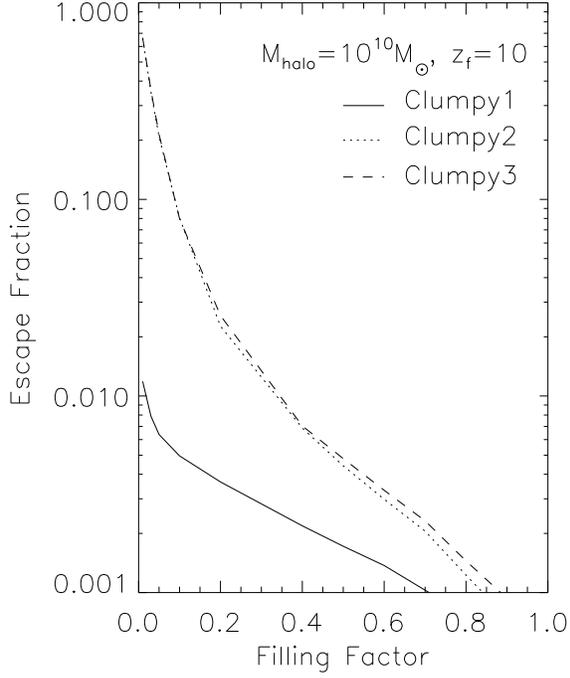}{6in}{90}{65}{65}{20}{100}}
\caption{Escape fractions of stellar ionizing photons from a disk within a
$10^{10}M_\odot$ halo formed at $z_f=10$.  The curves show different clumpy
models.  The volume filling factor refers to either the ionizing
emissivity, the clumps, or both, depending on the case.  Clumpy1: clumpy
emissivity, smooth interstellar medium (ISM); Clumpy2: smooth emissivity,
clumpy ISM; Clumpy3: uncorrelated clumpy emissivity and clumpy ISM.  }
\label{fig:10}
\end{figure}

\begin{figure}[t]
\centerline{\epsfysize=5.7in\epsffile{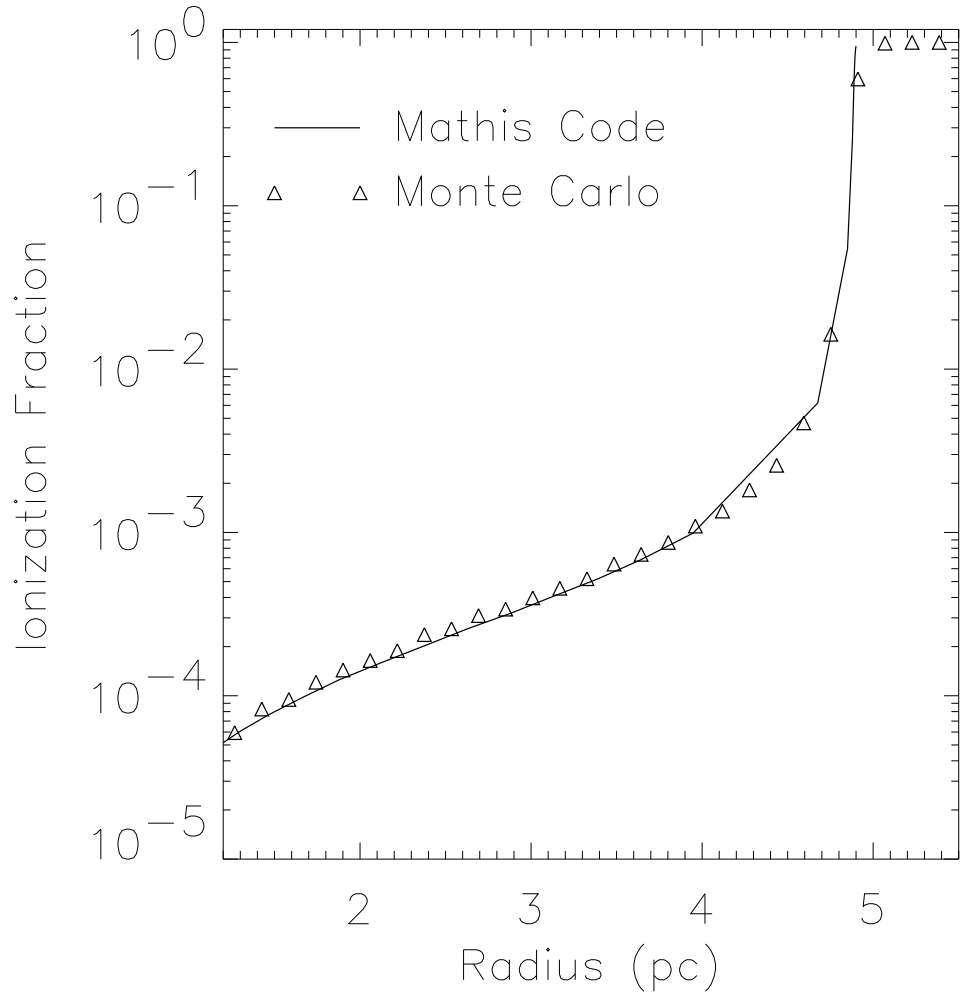 }}
\caption{Comparison of Monte Carlo photoionization with the code
of Mathis (1985).
Central source of ionizing luminosity $Q(H^0) = 4.26\times 10^{49}$s$^{-1}$
ionizes a region of constant number density $n_H = 100$cm$^{-3}$.}
\label{fig:11}
\end{figure}

\begin{figure}[t]
\centerline{\epsfysize=5.7in\epsffile{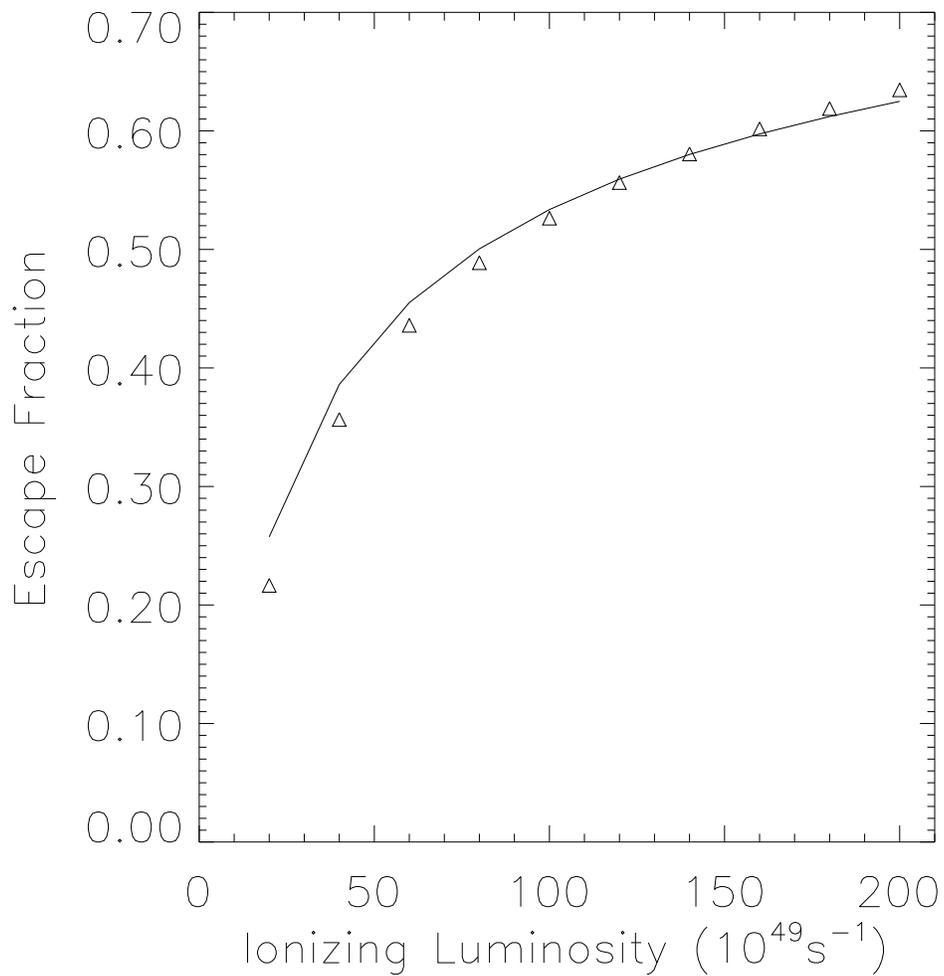 }}
\caption{Comparison of escape fractions calculated with Monte Carlo
photoionization code and the Str\"omgren volume approach of Dove \& Shull (1994).
Escape fractions are calculated for sources of different ionizing luminosity
that are embedded within a three component ``Dickey-Lockman'' disk.  Solid
curve is the analytic results from Dove \& Shull (1994), triangles are Monte
Carlo photoionization calculations.}
\label{fig:12}
\end{figure}

\begin{figure}[t]
\centerline{\plotfiddle{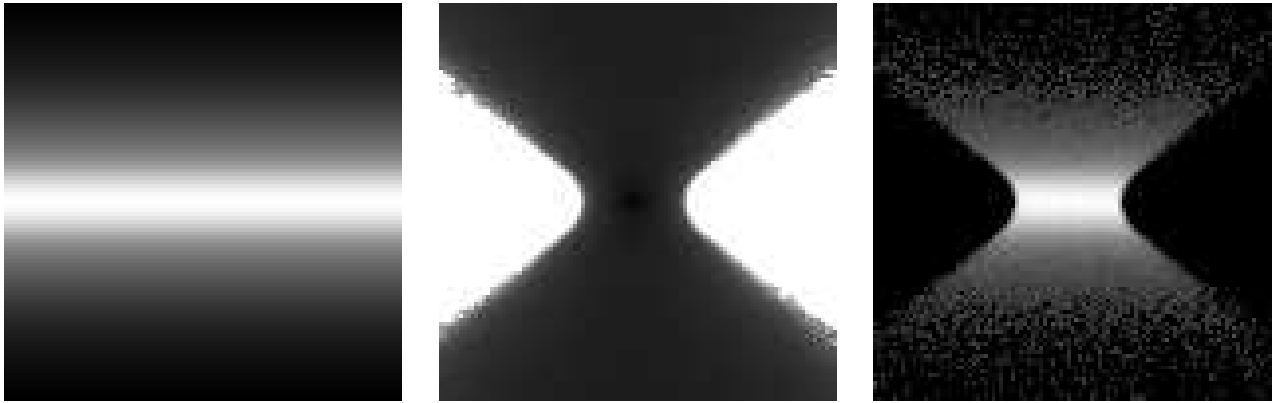}{6in}{-90}{65}{65}{-500}{400}}
\caption{Slices through our Monte Carlo photoionization model for the
$Q=2\times 10^{50}$s$^{-1}$ simulation of Figure~12.  
Left panel shows the total density, middle panel illustrates the ionization
fraction (white is neutral), and the right panel shows locations of
absorbed photons.  The non-zero HI opacity within the Str\"omgren volume
traps some ionizing photons and explains why our escape fractions are
slightly smaller than those of Dove \& Shull (1994).}  \label{fig:8}
\label{fig:13}
\end{figure} 
\end{document}